# HE-LHC BEAM- PARAMETERS, OPTICS AND BEAM-DYNAMICS ISSUES

O. Brüning, O. Dominguez, S. Myers, L. Rossi, E. Todesco, F. Zimmermann

*CERN, Geneva, Switzerland*

*Abstract*

The Higher-Energy LHC (HE-LHC) should collide two proton beams of 16.5-TeV energy, circulating in the LHC tunnel. We discuss the main parameter choices, as well as some optics and beam dynamics issues, in particular the time evolution of emittances, beam-beam tune shift and luminosity, with and without controlled emittance blow up, considering various constraints, and the quadrupole-magnet parameters for arcs and interaction regions.

## MAIN PARAMETERS

The HE-LHC beam energy of 16.5 TeV corresponds to a dipole magnet of about 20-T field (see Table 1). These values should be compared with the LHC design parameters of 7 TeV and 8.33 T. They assume an identical geometry and the same bending-magnet filling factor. It should be noted that the 20 T operational field level is the upper limit of a 16-20 T range being considered and must be understood as design target value. Only a thorough global optimization study can indicate the most convenient, or simply the possible, field strength for the main dipoles.

The target peak luminosity at 33 TeV c.m. energy is chosen as $2 \times 10^{34}$ cm$^{-2}$s$^{-1}$[1], i.e. equal to twice the LHC design luminosity. At this luminosity value the radiation effects in the interaction region (IR), e.g. for the final triplet magnets and the detectors, are similar to those for the High-Luminosity LHC (HL-LHC) at 7 TeV beam energy with a target peak luminosity of $5 \times 10^{34}$ cm$^{-2}$s$^{-1}$. The IR radiation sensitivity, related to the collisions, is taken to scale with the product of beam energy and luminosity. We assume that the IR solutions found for the HL-LHC will also suit the HE-LHC IR. The HL-LHC already pushes the requirements to near - or beyond - the present state of the art.

The interaction-point (IP) beta functions are set to values between 0.4 and 1.0 m, which is comparable to the 0.55 m of the LHC design, and larger than for the HL-LHC (where proposed values range between 7 and 30 cm). Differently from LHC, the HE-LHC IP beta functions and emittances may be unequal in the two transverse planes.

The normalized transverse emittances at the start of a physics store are assumed to be in the range 1.8-3.8 μm - possibly different in the horizontal and vertical plane - and, hence, similar to those of both the nominal and the present LHC.

A total number of 1404 bunches is considered, at 50 ns spacing, at slightly more than the LHC design bunch intensity. The smaller than nominal number of bunches limits the beam-screen heat load from synchrotron radiation and image currents, keeps the stored beam energy at 480 MJ, close to the 360 MJ design value of LHC, which is important for machine protection, and has the additional benefit that the electron cloud is more benign than for a bunch spacing of 25 ns. The HE-LHC will feature additional electron-cloud mitigation measures like coatings or distributed clearing electrodes. An alternative scenario with 2808 bunches per beam, at 25 ns spacing, could operate at half the bunch charge with half the transverse emittance, with the same stored beam energy. This scenario would, however, be more challenging for machine protection and collimation, due to the increased transverse energy density, and is also likely to give rise to stronger electron-cloud effects.

The arc-dipole coil aperture is taken to be 40 mm, which is the same value as the original design value of the Superconducting Super Collider (SSC) [before it was increased and the project ultimately cancelled]. For comparison, the LHC coil diameter is 56 mm.

Taking into account margins for beam tube and beam screen, the related beam half aperture is reduced from 20 mm for the LHC to 13 mm for the HE-LHC. This represents a reduction of about 30%. The arc maximum aperture is needed at injection. A reduced aperture is acceptable since the HE-LHC injection energy will be higher than for the LHC.

Specifically, the HE-LHC injection energy is assumed to be equal to, or higher than, 1 TeV. This energy is chosen to confine the HE-LHC energy ramp to a factor of not much more than 16-20, similar to the present LHC. The beam energy of the SPS, serving as LHC injector, does not exceed 450 GeV. For the HE-LHC a new injector with beam energy above 1 TeV will be required.

With the assumed number of bunches and peak luminosity, the maximum number of events per crossing comes out to be about 4 times the nominal LHC, or 76, which is below the peak pile up considered for the HL-LHC. In this estimate, the total inelastic cross section at 33 TeV c.m. energy is assumed to be similar to the one at 14 TeV, i.e. about 60 mbarn.

The longitudinal emittance damping time from synchrotron radiation can be computed to be 1 hour, which is to be compared with 13 h for the nominal LHC. The synchrotron radiation leads to a rapid shrinkage of all three emittances, which can be controlled by noise injection in order to stabilize the beam with regard to impedance-driven instabilities or the beam-beam interaction.

Table 1: Flat and round-beam HE-LHC parameters [1].

| | nominal LHC | HE-LHC | |
|---|---|---|---|
| beam energy [TeV] | 7 | 16.5 | |
| dipole field [T] | 8.33 | 20 | |
| dipole coil aperture [mm] | 56 | 40 | |
| beam half aperture [cm] | 2.2 (x), 1.8 (y) | 1.3 | |
| injection energy [TeV] | 0.45 | >1.0 | |
| #bunches | 2808 | 1404 | |
| bunch population [$10^{11}$] | 1.15 | 1.29 | 1.30 |
| initial transverse normalized emittance [μm] | 3.75 | 3.75 (x), 1.84 (y) | 2.59 (x & y) |
| initial longitudinal emittance [eVs] | 2.5 | 4.0 | |
| number of IPs contributing to tune shift | 3 | 2 | |
| initial total beam-beam tune shift | 0.01 | 0.01 (x & y) | |
| maximum total beam-beam tune shift | 0.01 | 0.01 | |
| beam circulating current [A] | 0.584 | 0.328 | |
| RF voltage [MV] | 16 | 32 | |
| rms bunch length [cm] | 7.55 | 6.5 | |
| rms momentum spread [$10^{-4}$] | 1.13 | 0.9 | |
| IP beta function [m] | 0.55 | 1 (x), 0.43 (y) | 0.6 (x & y) |
| initial rms IP spot size [μm] | 16.7 | 14.6 (x), 6.3 (y) | 9.4 (x & y) |
| full crossing angle [μrad] | 285 (9.5 $\sigma_{x,y}$) | 175 (12 $\sigma_{x0}$) | 188.1 (12 $\sigma_{x0}$) |
| Piwinski angle | 0.65 | 0.39 | 0.65 |
| geometric luminosity loss from crossing | 0.84 | 0.93 | 0.84 |
| stored beam energy [MJ] | 362 | 478.5 | 480.7 |
| SR power per ring [kW] | 3.6 | 65.7 | 66.0 |
| arc SR heat load dW/ds [W/m/aperture] | 0.17 | 2.8 | 2.8 |
| energy loss per turn [keV] | 6.7 | 201.3 | |
| critical photon energy [eV] | 44 | 575 | |
| photon flux [$10^{17}$/m/s] | 1.0 | 1.3 | |
| longitudinal SR emittance damping time [h] | 12.9 | 0.98 | |
| horizontal SR emittance damping time [h] | 25.8 | 1.97 | |
| initial longitudinal IBS emittance rise time [h] | 61 | 64 | ~68 |
| initial horizontal IBS emittance rise time [h] | 80 | ~80 | ~60 |
| initial vertical IBS emittance rise time [h] | ~400 | ~400 | ~300 |
| events per crossing | 19 | 76 | |
| initial luminosity [$10^{34}$ cm$^{-2}$s$^{-1}$] | 1.0 | 2.0 | |
| peak luminosity [$10^{34}$ cm$^{-2}$s$^{-1}$] | 1.0 | 2.0 | |
| beam lifetime due to $p$ consumption [h] | 46 | 12.6 | |
| optimum run time $t_r$ [h] | 15.2 | 10.4 | |
| integrated luminosity after $t_r$ [fb$^{-1}$] | 0.41 | 0.50 | 0.51 |
| opt. av. int. luminosity per day [fb$^{-1}$] | 0.47 | 0.78 | 0.79 |

The emittance shrinkage allows for a natural and easy way of leveling the luminosity or the beam-beam tune shift, simply by controlling the amount of noise injected to blow up the beam, without any changes of optics, orbit or crab-cavity voltage.

The synchrotron-radiation heat load is approximately 2.8 W/m/aperture, significantly higher than the value of 0.17 W/m/aperture for the nominal LHC, and slightly above the maximum local cooling available with the present beam-screen capillaries. The total synchrotron radiation power per beam is 66 kW, almost a factor 20 higher than the 3.6 kW for the nominal LHC, but still close to the capacity limit of the existing LHC cryogenic plants [1,2].

The 400-MHz RF voltage is taken to be 32 MV, which is twice the nominal value of 16 MV. This value had been chosen to keep the synchrotron tune approximately the same as for the present LHC (which might be important for beam and particle stability). A value of 16 MV as for the nominal LHC is also possible, however [3]. In order to maintain Landau damping the longitudinal emittance ($4\pi\sigma_z\sigma_E$) is increased with the square root of the beam energy [4], to about 4 eVs at 16.5 TeV, starting from a value of 2.5 eVs at 7 TeV. Together with the assumed RF voltage this yields an rms bunch length of 6.5 cm not much shorter than the nominal value at 7 TeV of 7.55 cm. With 16 MV RF voltage, and for the same longitudinal emittance, the rms bunch length would be 8.0 cm.

The beam lifetime due to proton consumption is about 13 h, to be compared with 46 h for the nominal LHC and about 10 h for the HL-LHC. For both energies a total cross section of 100 mbarn is considered. The optimum run time is about 10 h assuming a 5-h turnaround time. This is somewhat shorter than the 15-h run time for the nominal LHC, due to the higher luminosity. The optimum average luminosity per day is about 0.8 fb$^{-1}$, or some 60% larger than an optimistic value of 0.5 fb$^{-1}$ for the nominal LHC.

The maximum total beam-beam tune shift for 2 IPs varies between 0.01 and 0.03. The maximum value can be restricted through transverse emittance control by noise injection. Without such external noise, the transverse emittance would result from the interplay of synchrotron radiation damping, intrabeam scattering, and the beam-beam interaction, which is a topic to be further investigated (see also [5]).

Both flat-beam and round-beam HE-LHC scenarios exist, as is illustrated in Table 2. The two scenarios promise similar luminosity performance.

The crossing angle for the nominal LHC corresponds to a separation of 9.5$\sigma_{x,y}$ at the parasitic long-range collision points around the IP. For the HE-LHC the crossing angles chosen provide an initial separation of 12$\sigma_{x0}$ at the close-by parasitic encounters and an even larger normalized separation after emittance shrinkage. Therefore, long-range beam-beam effects should not be important for the HE-LHC. Table 1 presents a more complete list of HE-LHC parameters [1].

Table 2: Flat & round-beam scenarios for the HE-LHC.

|  | nominal (round) | HE-LHC Flat | round |
|---|---|---|---|
| $\gamma\varepsilon$ (μm) | 3.75 | 3.75 (x), 1.84 (y) | 2.59 (x&y) |
| $\beta^*$ (m) | 0.55 | 1 (x), 0.43 (y) | 0.6 (x&y) |
| $\sigma^*$ [μm] | 16.7 | 14.6 (x), 6.3 (y) | 9.4 (x&y) |
| $\theta_c$ [μrad] | 285 | 175 | 188 |

## LUMINOSITY TIME EVOLUTION

Figure 1 shows the emittance evolution, for both flat and round beams, during a physics store with and without controlled emittance blow up. The luminosity evolution for the case with controlled blow up, in order to limit the total beam-beam tune shift to a value of 0.01, is illustrated in Fig. 2, which also demonstrates the equivalent performance of flat-beam and round-beam collisions. Figure 3 presents the time evolution of the corresponding integrated luminosities.

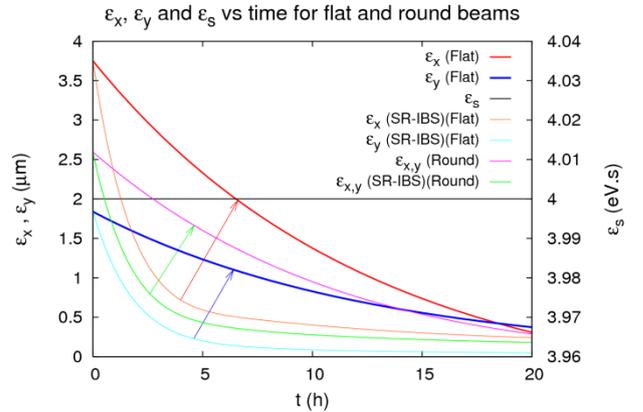

Figure 1: Evolution of the HE-LHC emittances, for flat and round beams, during a physics store with controlled blow up and constant longitudinal emittance of 4 eVs plus constant crossing angle (the thicker lines at the top), and the natural transverse emittance evolution due to radiation damping and IBS only (the thinner lines at the bottom) – still for constant longitudinal emittance and constant crossing angle, which might lead to excessive tune shifts.

What happens if we drop the constraint $\Delta Q_{tot} \leq 0.01$? This question is legitimate as the LHC has already reached a value of $\Delta Q_{tot} \sim 0.02$ (about twice the design value) without evidence for a beam-beam limit, and since LHC strong-strong beam-beam simulations by K. Ohmi, e.g. in [5], predict the LHC beam-beam limit at $\Delta Q_{tot} > 0.03$.

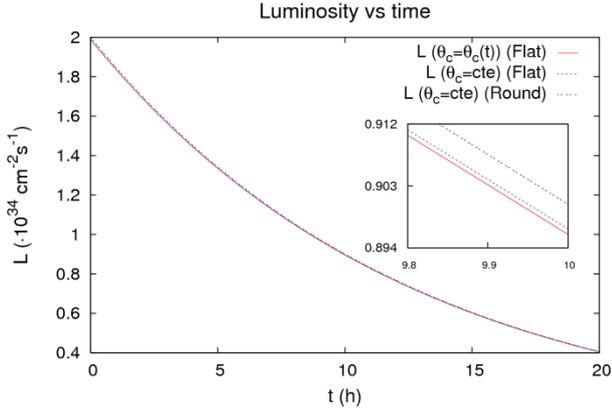

Figure 2: Time evolution of the HE-LHC luminosity, for both flat and round beams, including emittance variation with controlled blow up and proton burn off. Curves with constant or varying crossing angle lie on top of each other if the beam-beam tune shift is kept constant as assumed here.

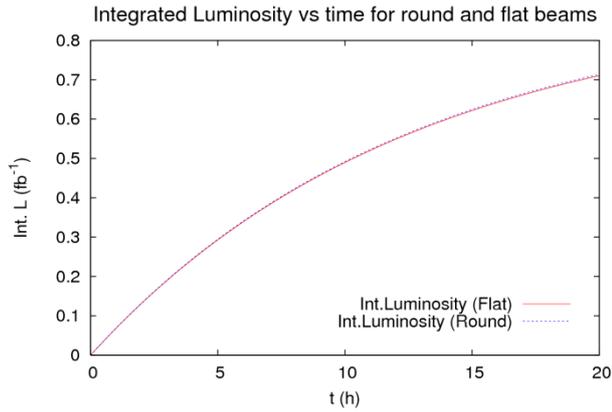

Figure 3: Time evolution of the HE-LHC integrated luminosity, for both flat and round beams, during a physics store including emittance variation with controlled blow up, keeping $\Delta Q_{tot} \leq 0.01$, and proton burn off.

Figure 4 shows the predicted tune shifts as a function of time during a physics store in the presence of synchrotron radiation damping and proton burn off, without any transverse emittance blow up, for flat and round beams, respectively. With flat beams the peak tune shift exceeds 0.03, with round beams it is about 0.02. In view of this difference, the round-beam option appears to be more conservative, with more than 30% lower beam-beam tune shift.

Figures 5 and 6 present the corresponding time evolutions of instantaneous and integrated luminosity, respectively, again with synchrotron-radiation and proton burn off, but without any controlled blow up. The gain in integrated luminosity of about 10% for the flat-beam case is much smaller than the increase in the peak beam-beam tune shift.

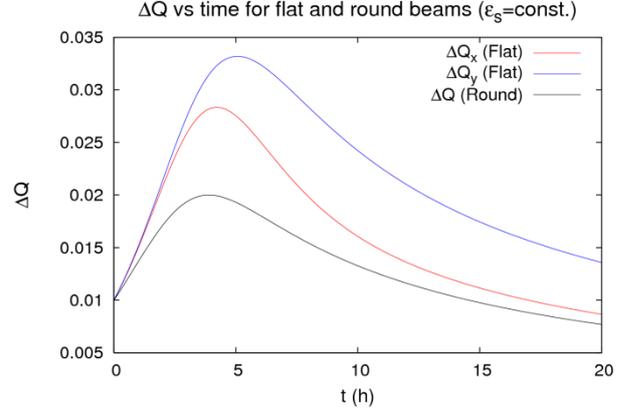

Figure 4: Time evolution of the HE-LHC tune shifts, for flat and round beams during a physics store including SR emittance shrinkage *without* controlled transverse blow up, and including proton burn off.

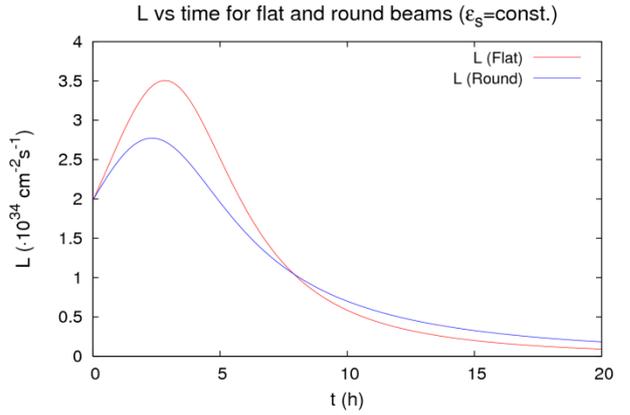

Figure 5: Time evolution of the HE-LHC instantaneous luminosity, for both flat and round beams, including SR emittance shrinkage and proton burn off, without controlled transverse blow up.

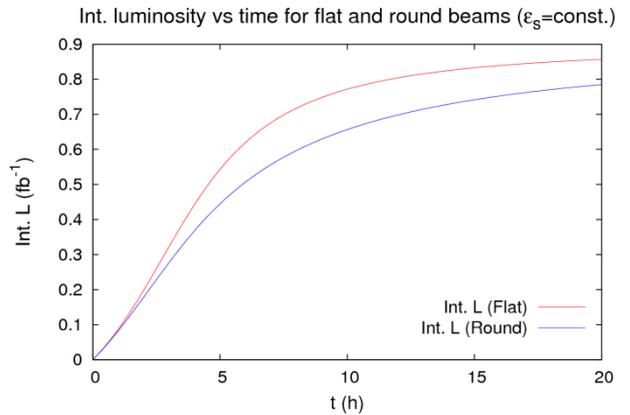

Figure 6: Time evolution of the HE-LHC integrated luminosity, for both flat and round beams, including SR emittance shrinkage and proton burn off, without controlled transverse blow up.

The sensitivity of the integrated luminosity to some of the assumptions has been investigated. For the baseline

HE-LHC we have 0.8 fb$^{-1}$/day as optimum average luminosity value (without any downtime and 100% availability). Without longitudinal blow up the average luminosity would be 5-20% lower, and without transverse blow up 10-20% higher. Another 25% increase of the average luminosity could be obtained, for round beams, with the ultimate bunch intensity of $1.7 \times 10^{11}$ protons, along with a larger initial transverse normalized emittance of 3.6 μm, and β* ~ 0.8 m (instead of 0.6 m).

## QUADRUPOLE MAGNETS

How do the interaction-region magnets scale with energy and β*? Can one hope to get a β* of 0.5-0.6 m, similar to the nominal LHC, at 2.36 times higher beam energy? Figure 7 illustrates the interdependence of the peak beta function in the final quadrupoles, the quadrupole gradient, the magnetic field at a radius of 16.5σ plus 11 mm (margin for beam screen, orbit and alignment errors, etc), and the IP beta function for 7 TeV beam energy, considering a triplet configuration [6]. Figure 8 converts Fig. 7 to 16.5 TeV beam energy, where the gradient scales with the beam energy, and the beam size with the square root of the energy and with the square root of the normalized emittance. For example, in order to achieve β*=0.55m at 16.5 TeV, a gradient of 400 T/m results in a peak beta function of about 4 km. With a normalized emittance γε=2.64 μm, the full beam aperture needed (33 σ) is about 26 mm. This point is indicated by a blue star in the parameter plane of Fig. 8.

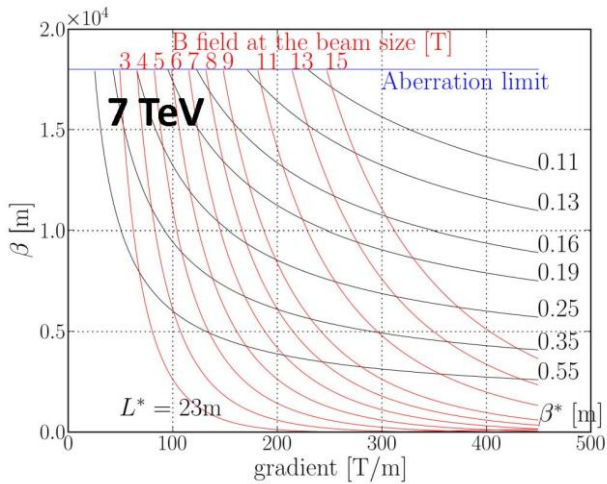

Figure 7: Peak beta function as a function of quadrupole gradient (horizontal axis), β* (red curves) and magnetic field at 16.5σ+11 mm (black curves) for 7 TeV beam energy [6].

For the arc quadrupoles we assume a full coil aperture of 40 mm as for the arc dipole magnets. If the length of the arc quadrupoles is the same as in the present LHC, their gradient must increase in proportion to the beam energy, from 223 T/m at 7 TeV to 526 T/m at 16.5 TeV. These scaled arc quadrupoles would then be more demanding the IR quadrupoles. Most probably the gradient of the arc quadrupoles needs to be lowered, or their aperture reduced. Aperture reduction is more attractive since lowering the gradient will probably lower the dipole field margin or the operating field and, in consequence, the beam energy. Clearly this point needs a thorough investigation.

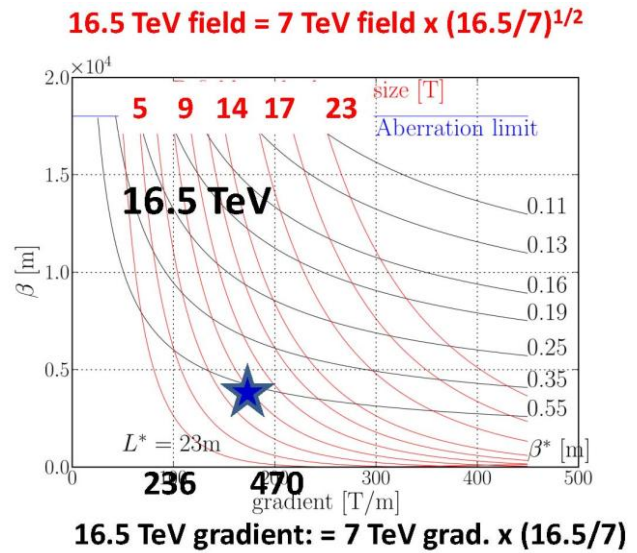

Figure 8: Peak beta function as a function of quadrupole gradient (horizontal axis), β* (red curves) and magnetic field at 16.5σ+11 mm (black curves), obtained by extrapolating Fig.8 to 16.5 TeV beam energy [the scaled values for 16.5 TeV are printed in bold face on top].

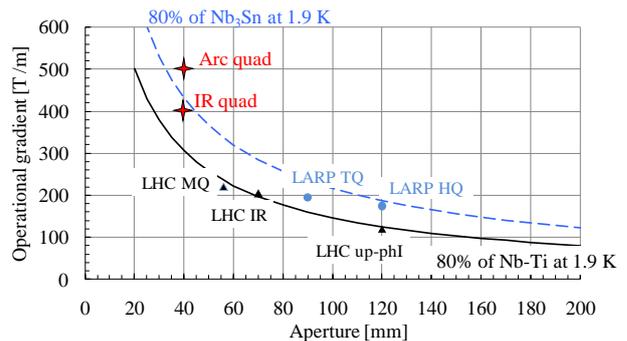

Figure 9: Operational gradient as a function of coil aperture for LHC and US-LARP quadrupoles (markers), scaling laws for limits in Nb.Ti and Nb3Sn (solid curves) [7], and expected values for HE LHC arc and IR (stars).

Figure 9 illustrates the location of the HE-LHC quadrupoles with respect to the LHC and LARP quadrupoles in the gradient-aperture plot [7]. The HE-LHC IR quadrupole still looks feasible with Nb$_3$Sn. However, a 40 mm aperture quadrupole for the arcs with 500 T/m is above the possibilities of Nb$_3$Sn. We would propose to aim for 400 T/m, which is at the limit of Nb$_3$Sn, and to compensate this lower gradient by a 20% increase in arc-quadrupole length (from 3.1 to 3.6 m). The integrated quadrupole strength required in the arcs also depends on the optical cell length, which sets the values for the beta functions. One should consider the possibility of changing the cell length with respect to LHC in order to find at a better optimization between long cell length, implying less quadrupoles and more space for bending, and short cell length, yielding lower beta functions and smaller aperture in the arcs.

## MISCELLANEOUS OPEN ISSUES

A larger number of points, mostly related to the higher beam energy, are outstanding and require further studies, e.g.
- the required cleaning efficiency assuming nominal quench levels;
- estimates of expected local radiation levels and implications for the dog-leg magnets in the cleaning insertions, and for the TAS and TAN designs;
- the required power converter tracking accuracy and potential implications if the HL-LHC features ca. 30-40 independent sectors (higher stored electro-magnetic energy in the magnets);
- stronger kicker elements for beam disposal (doubling the number of 15 dump kicker elements will have an impact on space and reliability), for beam diagnostics [tune measurements] and for generating large oscillation amplitudes [AC dipole, aperture kicker]), injection kickers & beam transfer with higher injection energy;
- beam diagnostics limits, e.g. for the use of beam screens and wire scanners;
- a closer inspection of the loss of longitudinal Landau damping; and the associated trade-off between bunch length and longitudinal impedance;
- persistent-current effects and field quality at injection which might, or might not, constrain the minimum injection energy required;
- the best gradient/aperture/length parameter set for the arc quadrupoles; and
- the use of crab cavities for HE-LHC: are crab cavities needed for HE-LHC? And/or could they be useful (e.g. suppose they are inherited from the HL-LHC)?

## SUMMARY

The proposed key parameters for the Higher-Energy LHC have been reviewed and justified. A few beam-dynamics and optics issues have been highlighted, such as the fast radiation damping, the resulting potentially high beam-beam tune shifts, the implied need for transverse and longitudinal emittance control, and the requirements for quadrupoles in the arcs and in the IRs. The realization of the HE-LHC project will depend on the future availability and affordability of high-field dipole magnets.

## ACKNOWLEDGEMENTS

We thank R. Assmann, R. Bailey, R. De Maria, G. De Rijk, B. Goddard, M. Jimenez, P. McIntyre, K. Ohmi, and L. Tavian for inspiring and helpful discussions.

The perfect workshop organization by Nicholas Sammut and Merethe Morer-Olafsen has been much appreciated.